\documentclass[ISTS]{tjsass} 

\usepackage[dvipdfmx]{graphicx}
\RequirePackage{multicol}
\RequirePackage{amsmath}
\RequirePackage[varg]{txfonts}
\RequirePackage{bm}
\RequirePackage{array}
\RequirePackage{color}
\RequirePackage{tjsasscite}
\usepackage[hang]{footmisc}
\newcommand{\bhline}[1]{\noalign{\hrule height #1}}

\usepackage{url}
\usepackage{amssymb}
\usepackage{siunitx}

\pubyear{2020}
\bookvolume{17}
\bookissue{20}
\titleheadertrue
\receiveddate{March 8th, 2023}%

\title{Thermophysical Model Development for Hera Mission to Simulate Non-Gravitational Acceleration on Binary Asteroid}
\subtitle{}
\author{
    \NAME{Masanori}{Kanamaru},\thanksNum{1)}\CorresAuthor{kanamaru@eps.s.u-tokyo.ac.jp}
    \NAME{Tatsuaki}{Okada},   \thanksNum{2)}
    \NAME{Hiroki}{Senshu},    \thanksNum{3)}
    \NAME{Hirohide}{Demura},  \thanksNum{4)}
    \NAME{Naru}{Hirata},      \thanksNum{4)}
    \NAME{Yuto}{Horikawa}     \thanksNum{5)} and
    \NAME{Giacomo}{Tommei}    \thanksNum{6)}
}
\thanksOrg{The University of Tokyo, Bunkyo, Japan}
\thanksOrg{Institute of Space and Astronautical Science, JAXA, Sagamihara, Japan}
\thanksOrg{Planetary Exploration Research Center, Chiba Institute of Technology, Chiba, Japan}
\thanksOrg{The University of Aizu, Aizu-Wakamatsu, Japan}
\thanksOrg{Osaka University, Toyonaka, Japan}
\thanksOrg{The University of Pisa, Pisa, Italy}
\begin{abstract}

The surface temperature of an asteroid is fundamental information for the design of an exploration mission and the interpretation of scientific observations. In addition, the thermal radiation of the asteroid causes a non-gravitational acceleration that induces secular changes in its orbit and spin. We have been developing a numerical calculation library for simulating the dynamics and thermophysics of asteroids. The asteroid dynamical simulator, \texttt{Astroshaper}, can calculate the temperature distribution based on a 3-dimensional shape model of an asteroid and predict the non-gravitational acceleration. In recent years, asteroid exploration missions such as Hayabusa2 and Hera have been equipped with thermal infrared imagers. The asteroid thermography can provide the thermal properties of the surface material of the target bodies. The functionality of thermophysical modeling in \texttt{Astroshaper} contributes to simulating the thermal environment on the asteroids, estimating the thermal properties, and predicting the dynamical evolution controlled by the non-gravitational effects.

\end{abstract}
\begin{document}
\maketitle
\renewcommand{\thefootnote}{\roman{footnote}}


\section*{Nomenclature}


\vbox{\noindent\setlength{\tabcolsep}{0mm}%
\begin{tabular}{p{15mm}cp{60mm}} %
    \hfil $A_{\rm B}$     \hfil & :\hspace{4mm} & Albedo at visible wavelength \\
    \hfil $A_{\rm TH}$    \hfil & :\hspace{4mm} & Albedo at thermal radiation wavelength \\
    \hfil $a$             \hfil & :\hspace{4mm} & Area of a facet, \si{m^2} \\
    \hfil $C_p$           \hfil & :\hspace{4mm} & Heat capacity at constant pressure, \si{J/kg/K} \\
    \hfil $c_0$           \hfil & :\hspace{4mm} & Speed of light in vacuum, \si{m/s} \\
    \hfil $d\bm{f}$       \hfil & :\hspace{4mm} & Thermal force on a facet, \si{N} \\
    \hfil $E$             \hfil & :\hspace{4mm} & Sum of emittance of scattered light and thermal radiation from a facet, \si{W/m^2} \\
    \hfil $E_{\rm cons}$  \hfil & :\hspace{4mm} & $E_{\rm out} \, / \, E_{\rm in}$ \\
    \hfil $E_{\rm in}$    \hfil & :\hspace{4mm} & Energy incident on an asteroid, \si{W} \\
    \hfil $E_{\rm out}$   \hfil & :\hspace{4mm} & Energy emitted from an asteroid, \si{W} \\
    \hfil $F_{\rm rad}$   \hfil & :\hspace{4mm} & Energy flux by thermal radiation from surrounding facets, \si{W/m^2} \\
    \hfil $F_{\rm scat}$  \hfil & :\hspace{4mm} & Energy flux by scattered light from surrounding facets, \si{W/m^2} \\
    \hfil $F_{\rm sun}$   \hfil & :\hspace{4mm} & Energy flux by direct sunlight, \si{W/m^2} \\
    \hfil $F_{\rm total}$ \hfil & :\hspace{4mm} & Total energy flux into a facet, \si{W/m^2} \\
    \hfil $f$             \hfil & :\hspace{4mm} & View factor between two facets \\
    \hfil $k$             \hfil & :\hspace{4mm} & Thermal conductivity, \si{W/m/K} \\
    \hfil $\bm{\hat n}$   \hfil & :\hspace{4mm} & Normal vector of a facet \\
    \hfil $\bm{r}$        \hfil & :\hspace{4mm} & Position vector, \si{m} \\
    \hfil $T$             \hfil & :\hspace{4mm} & Temperature, \si{K} \\
    \hfil $t$             \hfil & :\hspace{4mm} & Time, \si{s} \\
    \hfil $z$             \hfil & :\hspace{4mm} & Depth, m \\
\end{tabular}}

\vbox{\noindent\setlength{\tabcolsep}{0mm}%
\begin{tabular}{p{15mm}cp{60mm}} %
    \hfil $\alpha$        \hfil & :\hspace{4mm} & Thermal force on an asteroid, \si{N} \\
    \hfil $\Gamma$        \hfil & :\hspace{4mm} & Thermal inertia, \si{J \cdot m^{-2} \cdot K^{-1} \cdot s^{-0.5}} (tiu) \\
    \hfil $\varepsilon$   \hfil & :\hspace{4mm} & Emissivity \\
    \hfil $\theta$        \hfil & :\hspace{4mm} & Tilt angle of a facet \\
    \hfil $\rho$          \hfil & :\hspace{4mm} & Density, \si{kg/m^3} \\
    \hfil $\sigma$        \hfil & :\hspace{4mm} & Stefan--Boltzmann constant, \si{W/m^2/K^4} \\
    \hfil $\tau$          \hfil & :\hspace{4mm} & YORP torque on an asteroid, \si{N \cdot m} \\
\end{tabular}}

\noindent{Subscripts}

\vbox{\noindent\setlength{\tabcolsep}{0mm}%
\begin{tabular}{p{15mm}cp{60mm}}
    \hfil $\rm Didy$      \hfil & :\hspace{4mm} & Didymos \\
    \hfil $\rm Dimo$      \hfil & :\hspace{4mm} & Dimorphos, the satellite of Didymos \\
    \hfil $i$             \hfil & :\hspace{4mm} & Index of a facet of a shape model \\
    \hfil $j$             \hfil & :\hspace{4mm} & Index of a facet visible from facet $i$ \\
\end{tabular}}


\section{Introduction}

\subsection{Thermophysical modeling of an asteroid}

Thermophysical modeling (TPM) is a numerical simulation to obtain temperature distribution on the surface of an asteroid. TPM plays a vital role in a small-body mission's science and engineering aspects as follows.

\begin{itemize}
    \item TPM simulates the thermal environment around the asteroid that is critical for a proximity operation and a touch-down operation to the surface.
    \item It is possible to map the asteroid's thermal inertia and surface roughness by comparing TPM and thermal infrared spectroscopy or imaging \cite{Okada2020-fj, Shimaki2020-cd, Senshu2022-oh}.
    \item TPM can predict the non-gravitational acceleration on the asteroid induced by anisotropic thermal radiation. The changes in orbit and rotation of asteroids due to thermal radiation are known as the Yarkovsky and YORP effects, respectively \cite{Rubincam2000-tt, Bottke2006-dg}.
    \item The orbit evolution by the Yarkovsky effect is also important for assessing the risk of asteroid impact on Earth in planetary defense \cite{Giorgini2002-vk, Farnocchia2021-wm}.
    \item Changes in surface temperatures may cause material ejection from the asteroid and comet nuclei \cite{Rozitis2020-ab}.
    \item Thermal radiation pressure from the asteroid's surface affects the trajectory of the spacecraft or the ejecta particle in the vicinity of the asteroid \cite{McMahon2020-fl, Pedros-Faura2022-mn}.
    \item Thermal radiation causes a bias in the infrared spectra of the asteroid. To interpret the spectra at $\sim 3$ \textmu m or longer wavelength, removing this ``thermal tail"  is necessary\cite{Simon2019-ui}.
\end{itemize}

\subsection{Hera mission to explore a binary asteroid}

DART and Hera are planetary defense missions to a binary asteroid with a satellite \cite{Rivkin2021-ou, Michel2022-af}. The DART spacecraft successfully impacted Dimorphos, a satellite of the asteroid Didymos, in September 2022 \cite{Daly2023-zu}. The momentum transfer efficiency by the DART impact was estimated from the change in the mutual orbit period of the binary asteroid \cite{Thomas2023-iw, Cheng2023-ty}. The Hera spacecraft is scheduled to rendezvous with Didymos and Dimorphos in December 2026 to observe in detail the crater formed by the DART impact \cite{Michel2022-af}. Japan's team led by the Institute of Space and Astronautical Science (ISAS) is developing a thermal infrared imager (TIRI) onboard the Hera spacecraft. TIRI is the successor to the thermal infrared imager (TIR) on Hayabusa2, with higher sensitivity and resolution and six band filters for mid-infrared spectroscopy. Asteroid thermography by TIRI will provide us with the thermal inertia or density of the boulders and gravel that make up the target asteroids, which is essential for assessing the efficiency of the asteroid deflection experiment by DART.

\subsection{Development of thermophysical models for single/binary asteroids}

Several thermophysical models have been developed for single asteroids. One of the most elaborate models is the Advanced Thermophysical Model (ATPM), including the effect of small-scaled surface roughness \cite{Rozitis2011-ux}. We have been developing a numerical simulator for the dynamics and thermophysics of asteroids, \texttt{Astroshaper}. This simulator was originally developed for YORP prediction of asteroid Ryugu, a target asteroid of the Hayabusa2 mission \cite{Kanamaru2021-rx}. \texttt{Astroshaper} is being developed as an open-source project in the Julia programming language at GitHub\footnote{\texttt{Astroshaper} -- \textcolor{blue}{\url{https://github.com/Astroshaper}}}. We hereby report on the functionality of thermophysical modeling implemented in the \texttt{AsteroidThermoPhysicalModels.jl} package\footnote{\texttt{AsteroidThermoPhysicalModels.jl} -- \textcolor{blue}{\url{https://github.com/Astroshaper/AsteroidThermoPhysicalModels.jl}}}, one of the sub-modules of \texttt{Astroshaper}. Some sample codes for TPM simulation are also available in the repository of \texttt{Astroshaper-example}\footnote{\texttt{Astroshaper-example} -- \textcolor{blue}{\url{https://github.com/Astroshaper/Astroshaper-examples}}}. We have extended the capabilities of TPM for a single asteroid to apply to a binary asteroid for interpreting the TIRI imagery of Didymos and Dimorphos.


\section{TPM Functionality of \texttt{Astroshaper}}

The thermophysical model implemented in \texttt{AsteroidThermoPhysicalModels.jl} is based on a 3-dimensional shape model of an asteroid covered with a triangular mesh. As with other TPMs\cite{Rozitis2011-ux}, it can calculate the temperature distribution of the asteroid considering some fundamental thermophysical processes (See Table \ref{tbl: thermophysics}): the 3D shape of the asteroid, 1-dimensional heat conduction in the depth direction, shadowing by the local horizon (i.e., self-shadowing), and reabsorption of scattered light and thermal radiation by interfacing facets (i.e., self-heating).

\begin{table}[tbp]
    \centering
    \caption{Thermophysics implemented in \texttt{Astroshaper}.}
    \label{tbl: thermophysics}
    \begin{tabular}{lp{45mm}}
        \bhline{0.8pt}
        Asteroid 3D shape   & Yes. Triangular mesh models can be imported. \\
        Heat conduction     & Yes. 1D heat conduction in the depth direction is considered.\\
        Self-shadowing      & Yes. \\
        Self-heating        & Yes. Only single scattering is considered. \\
        Mutual-shadowing    & Yes. \\
        Mutual-heating      & Yes. \\
        Surface roughness   & Not yet implemented. \\
        \bhline{0.8pt}
    \end{tabular}
\end{table}%


\subsection{Heat conduction}

\begin{figure}[!t]%
    \centering
    \includegraphics[width=70mm,clip]{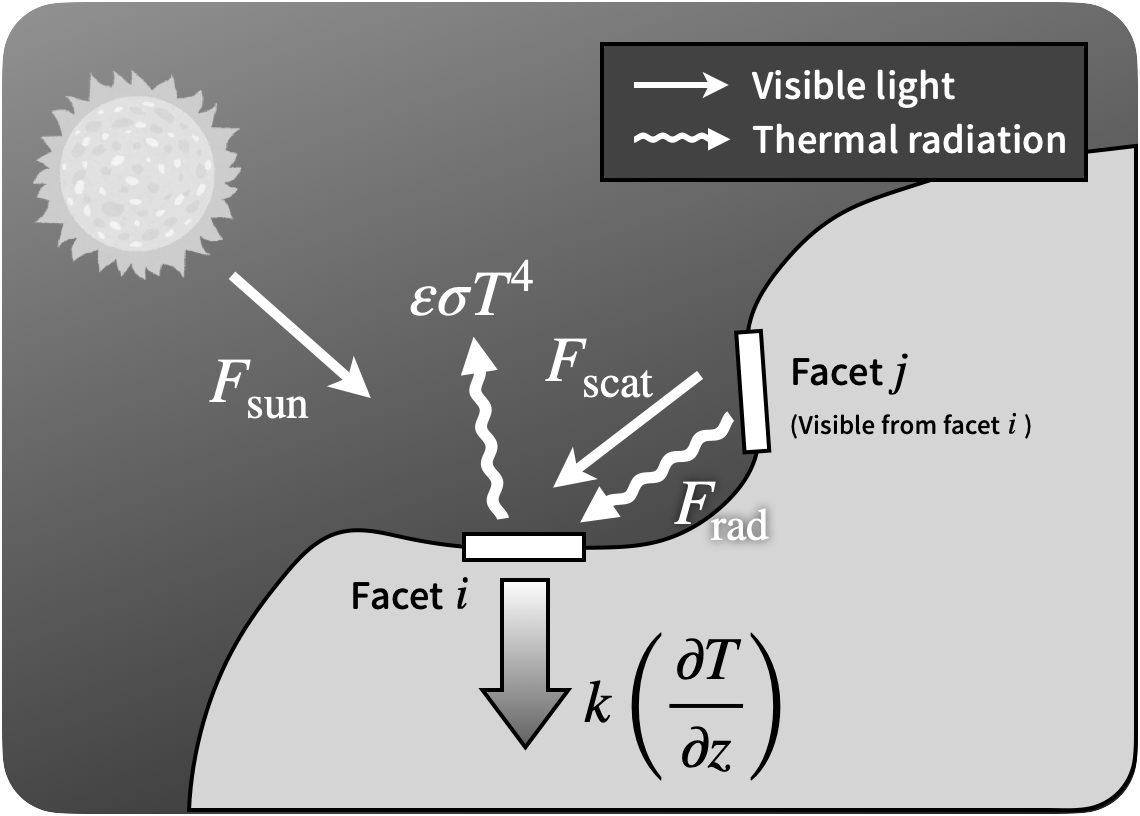}
    \caption{Basic thermophysical processes on an asteroid.}%
    \label{Figure_SingleTPM}%
\end{figure}

Our TPM code independently solves a 1-dimensional heat conduction equation on each shape model facet. Assuming that the thermal conductivity $k$ is constant regardless of depth $z$, the heat conduction equation becomes as follows.
    \begin{equation}
        \frac{\partial T}{\partial t} = \frac{k}{\rho C_p} \frac{\partial^2 T}{\partial z^2}
        \label{eq: heat conduction}
    \end{equation}
The boundary condition at the surface of the asteroid ($z=0$) is given by the balance of incident light to the facet, heat flux to the ground, and thermal radiation to space (See Fig. \ref{Figure_SingleTPM}).
    \begin{equation}
        F_{\rm total} + k \left( \frac{\partial T}{\partial z} \right)_{z=0} = \varepsilon \sigma T^4_{z=0}
        \label{eq: upper boundary condition}
    \end{equation}
where $F_{\rm total}$ is the total energy the facet absorbs at each time step.
    \begin{equation}
        F_{\rm total} = (1-A_{\rm B}) F_{\rm sun} + (1-A_{\rm B}) F_{\rm scat} + (1-A_{\rm TH}) F_{\rm rad}
        \label{eq: total flux}
    \end{equation}
The solar incident $F_{\rm sun}$ is an energy flux that considers the inclination of the facet concerning the sun's direction and the shadow of the surrounding facets. To consider the self-shadowing effect, $F_{\rm sun}$ is set to zero when the other facet blocks the solar ray. The facet exchanges the energy flux with other interfacing facets by reabsorbing the scattered light and thermal radiation. $F_{\rm scat}$ and $F_{\rm rad}$ are the energy fluxes from the interfacing facets to the facet in question in visible and thermal infrared wavelengths, respectively. In our model, single scattering is only considered. The additional flux due to multiple scattering is negligible for a low albedo body. The boundary condition of  insulation is given so that the temperature gradient is zero at the bottom cell.
    \begin{equation}
        \left( \frac{\partial T}{\partial z} \right)_{z \rightarrow \infty} = 0
        \label{eq: lower boundary condition}
    \end{equation}
Our TPM code solves the above equations by an explicit Euler scheme. The radiative boundary condition involving a nonlinear term at Eq. (\ref{eq: upper boundary condition}) is solved using the Newton-Raphson method. It is in the process of being implemented to allow users to select implicit and higher-order solvers.

\subsection{Non-gravitational force}

Non-gravitational perturbations on the asteroid can be calculated from the temperature distribution\cite{Rozitis2012-jg}. We assume that a facet of the shape model scatters and radiates isotropically (i.e., Lambertian scatterer and emitter). The total emittance of scattered light and thermal radiation emitted from facet $i$ is
    \begin{equation}
        E_i = A_{\rm B} F_{{\rm sun}, \, i} + A_{\rm B} F_{{\rm scat}, \, i} + A_{\rm TH} F_{{\rm rad}, \, i} + \varepsilon \sigma T_i^4
        \label{eq: emittance}
    \end{equation}
The force exerted by the photon pressure on the element can be expressed as follows.
    \begin{equation}
        d\bm{f}_i = - \frac{2 E_i a_i}{3 c_0} \bm{\hat n}_i + \sum_{j \, \in \,{\rm visible\,from\,facet}\, i} \frac{E_i a_i}{c_0} f_{i,j} \, \frac{\bm{r}_j - \bm{r}_i}{|\bm{r}_j - \bm{r}_i|}
        \label{eq: facet thermal force}
    \end{equation}
The first term is a force component normal to the surface element. The coefficient $-2/3$ is derived from the isotropic emittance. The second term represents the additional component due to the interaction with visible facets. The reabsorbed photons exert a force along the direction from facet $i$ to facet $j$ in proportion to the view factor $f_{i,j}$. The view factor from facet $i$ to facet $j$ refers to the fraction of absorption by facet $j$ to the emittance from facet $i$\cite{Rozitis2012-jg, Lagerros1998-rn}.
    \begin{equation}
        f_{i,j} = \frac{\cos{\theta_i} \cos{\theta_j}}{\pi \, |\bm{r}_j - \bm{r}_i|^2} a_j
        \label{eq: view factor}
    \end{equation}
where $\theta_i$ and $\theta_j$ are the angles between each normal vector and the line connecting the two facets, and $d_{i,j}$ denotes the distance between the two facets. The summation of Eq. (\ref{eq: facet thermal force}) should only be taken for facets seen from facet $i$. In our code, the visible facets from each facet are searched and stored before the TPM is performed.

The total force $\alpha$ and torque $\tau$ on the asteroid are obtained by integrating the thermal force over the entire surface.
    \begin{equation}
        \alpha = \sum_i \left( \frac{\bm{r}_i}{|\bm{r}_i|} \cdot d\bm{f}_i \right) \, \frac{\bm{r}_i}{|\bm{r}_i|}
        \label{eq: total thermal force}
    \end{equation}

    \begin{equation}
        \tau = \sum_i \bm{r}_i \times d\bm{f}_i
        \label{eq: total thermal torque}
    \end{equation}
The perturbation to the motion of the asteroid's center-of-mass causes the Yarkovsky drift in orbit, and the torque causes the YORP spin evolution.


\subsection{Binary and additional thermophysics}

\begin{figure}[!t]%
    \centering
    \includegraphics[width=70mm,clip]{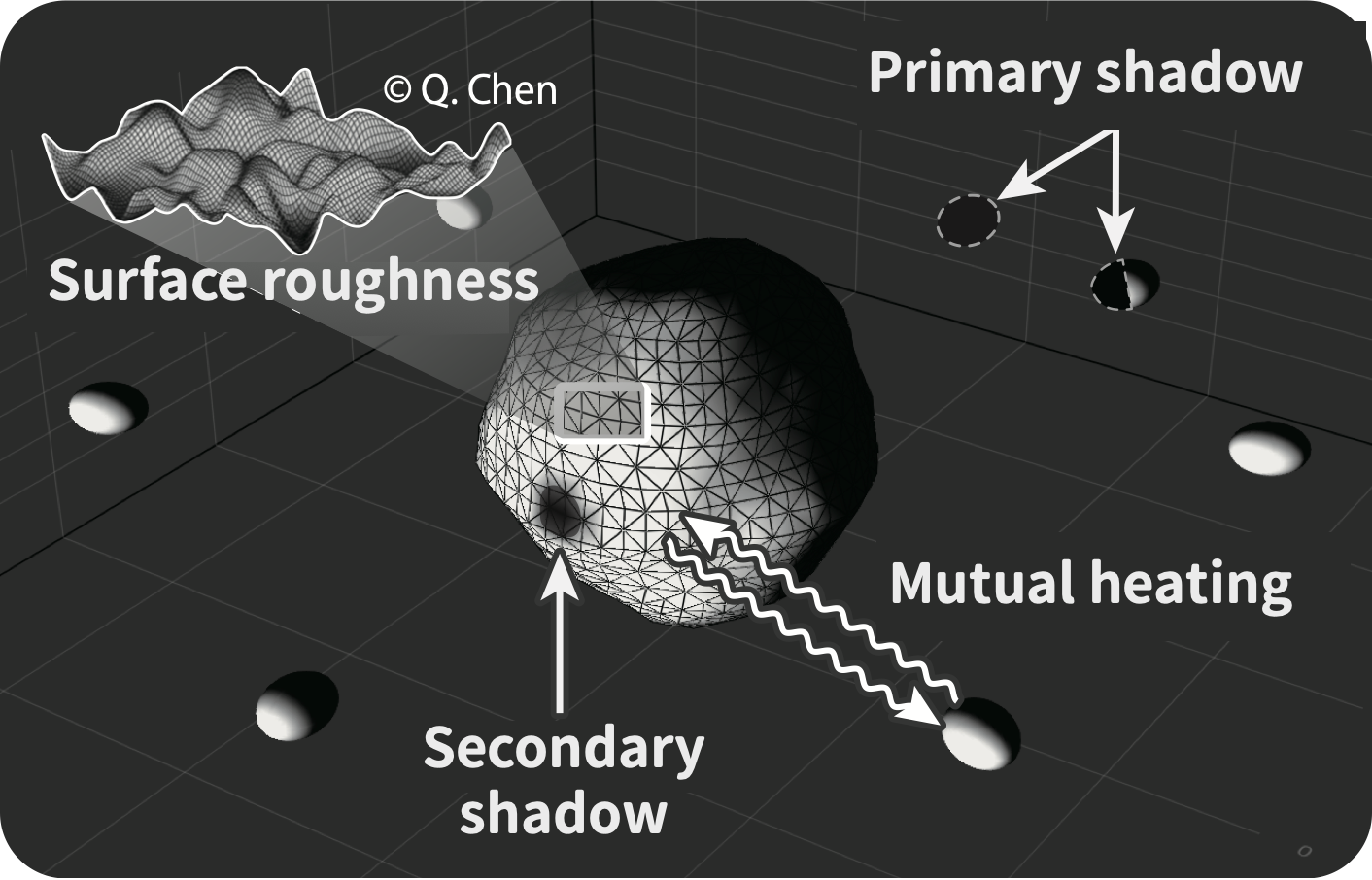}
    \caption{Thermophysics for a binary asteroid. The shape models of Didymos and Dimorphos based on ground-based observations are shown here.}%
    \label{Figure_BinaryTPM}%
\end{figure}

Some additional thermophysics must be considered for a binary asteroid, as in Fig. (\ref{Figure_BinaryTPM}). We utilized the functions of ray tracing for detecting local shadows on a single asteroid to simulate an eclipse by a pair of asteroids (i.e., mutual shadowing). Two types of eclipse events can occur: when the satellite's shadow falls on the primary asteroid and when the satellite enters the shadow of the primary. The primary and secondary asteroids exchange energy by thermal radiation and warm each other. This mutual heating effect is also implemented. The impact of the thermal infrared beaming by small-scaled surface roughness will be implemented in the future.

\section{TPM for Binary Asteroid Didymos and Dimorphos}



\subsection{Parameter setting}

We used the SPICE kernels and 3D shape models provided by the Hera mission for a thermophysical simulation of the binary asteroid Didymos and Dimorphos.\footnote{Shape models used in this study (Version: \texttt{v140\_20230731\_001}):
\begin{itemize}
    \item \texttt{g\_50677mm\_rad\_obj\_dida\_0000n00000\_v001.obj} for Didymos
    \item \texttt{g\_06650mm\_rad\_obj\_didb\_0000n00000\_v001.obj} for Dimorphos
\end{itemize}
Available at \textcolor{blue}{\url{https://s2e2.cosmos.esa.int/bitbucket/projects/SPICE_KERNELS/repos/hera/browse}}.} The shape models used in this study are based on ground-based observations before the DART impact experiment. It should be noted that the shape of Dimorphos is approximated by an ellipsoid.

A thermal inertia of $\Gamma = 403$ tiu was given, corresponding to a typical value for an S-type asteroid\cite{Delbo2015-zc}. Running TPM over tens of thermal cycles in advance is necessary to obtain a temperature distribution independent of initial conditions. In this study, TPM was performed for two months (from January 1st to March 1st, 2027) after temperatures of 0K were given at all facets of the shape models and all depth cells, corresponding to $\sim 627$ rotations for Didymos and $\sim 119$ mutual orbit cycles for Dimorphos. We confirmed that the calculation sufficiently converged in terms of the balance between the energy input and output on the surface of each asteroid, where $E_{\rm cons}$ was greater than 0.98 at the final time step. We used the simulated temperature data for 24 hours on March 1st, 2027, for the later analysis.

\subsection{Temperature map of the binary asteroid}

\begin{figure}[!t]%
    \centering
    \includegraphics[width=85mm,clip]{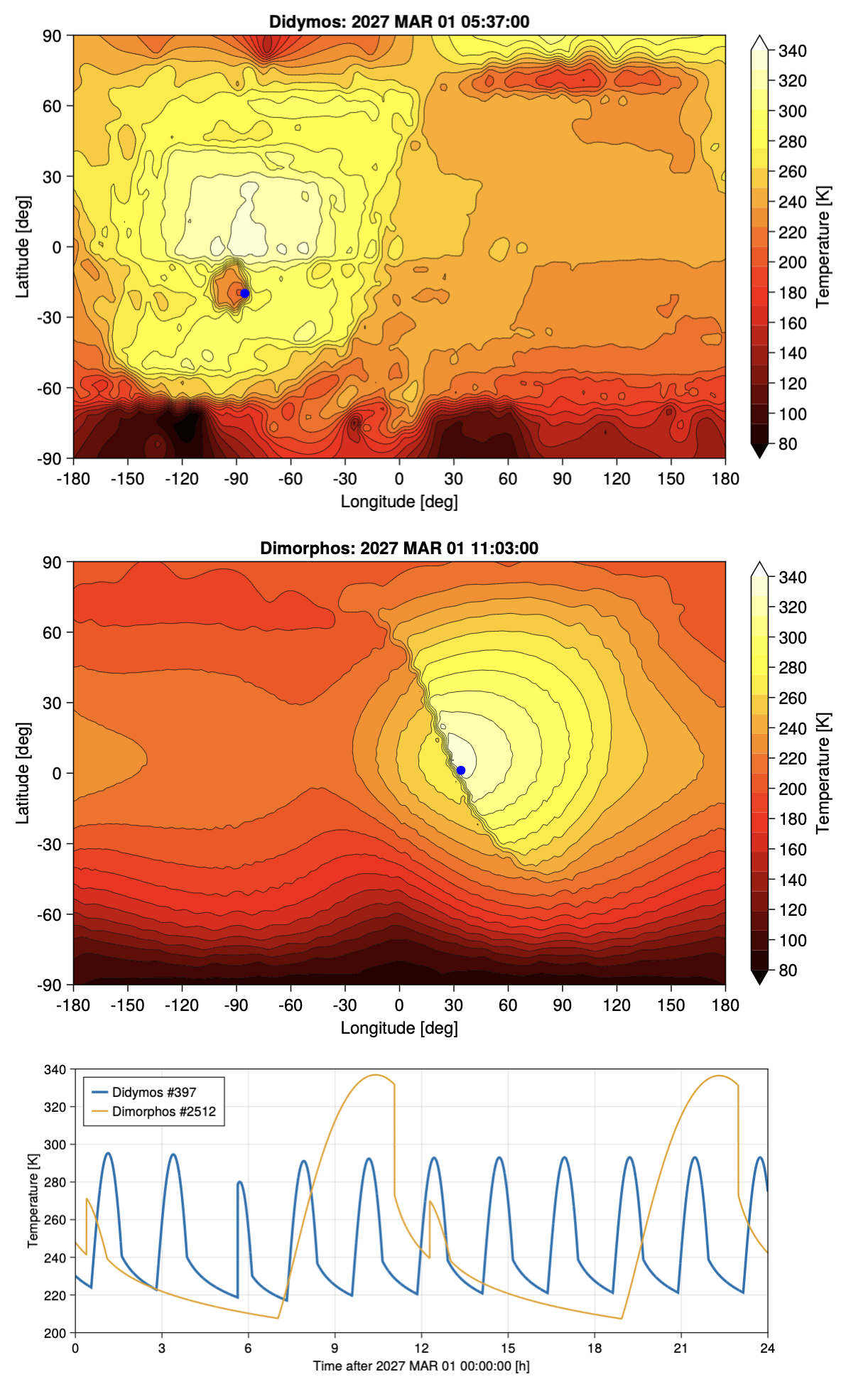}
    \caption{Temperature maps of Didymos (upper) and Dimorphos (middle), and temperature changes at the points indicated by the blue dots (lower).}%
    \label{Figure_TemperatureMaps}%
\end{figure}

The upper and middle panels of Fig. \ref{Figure_TemperatureMaps} show the temperature maps of Didymos and Dimorphos at the epochs of the mutual events, respectively. In the upper panel, Dimorphos cast the shadow around $(20^\circ \rm{S}, 90^\circ \rm{W})$ of Didymos at 5:37 a.m. After 5.96 hours or half of the orbit period of Dimorphos, one can observe Dimorphos hiding in the shadow of Didymos (middle panel). The lower panel shows the temperature changes over time at the points indicated by the blue dots on the above maps. It can be seen that rapid temperature drops of several tens of Ks occurred during the eclipse events. By observing the eclipse events in addition to diurnal thermal cycles, thermophysical properties corresponding to different depths can be investigated by TIRI. Because of the considerable uncertainty in the inclination of the mutual orbit, it will be turned out after Hera's rendezvous how frequently the eclipse events will occur.

\subsection{Non-Gravitational Effects on the binary asteroid}


Based on the above temperature distribution, we also calculated the thermal recoil force on each facet of the shape model. We integrated it over the surface to obtain non-gravitational force and torque on the binary asteroid. By averaging over several rotations, the torque components for rotational acceleration were estimated as $\tau_{\rm Didy} = 0.19 \,\, {\rm N \cdot m}$ for Didymos and $\tau_{\rm Dimo} = -1.1 \times 10^{-4} \,\, {\rm N \cdot m}$ for Dimorphos. It suggests that the rotation of Didymos is accelerating at the so-called YORP time scale of $4.1 \times 10^6$ years, that is, a time to double the rotation speed. On the other hand, the negative acceleration of Dimorphos decelerates its rotation at a time scale of $8.6 \times 10^4$ years, reducing the rotation speed by half.


\section{Discussion}


Generally, the resolution of a pre-arrival shape model is insufficient for YORP prediction sensitive to small-scale topography\cite{Statler2009-lx}. We must wait for Hera's rendezvous for a more precise prediction of YORP on Didymos and Dimorphos. The shape model of Dimorphos used in this study is an ellipsoid based on ground-based observations. The symmetrical shape should cancel out the thermal torque, but the asymmetry of the temperature distribution results in the non-zero torque. Cooling due to the eclipse is likely the cause of the negative acceleration on the satellite. The drastic temperature change may have the effects of expanding the mutual orbit of the binary asteroid and shortening its dynamical lifetime.


\section{Conclusion}

We hereby reported on the asteroid dynamical simulator, \texttt{Astroshaper}. We have developed a thermophysical simulation for the Hera mission applicable to a binary asteroid. This tool is expected to contribute to the operation planning of TIRI and investigate the dynamics of the binary asteroid controlled by the non-gravitational effects.

\section*{Acknowledgments}\label{Acknowledgments}

This study was supported by the JSPS KAKENHI No. JP17H06459 (the \textit{Aqua Planetology} project) and No. JP22J00435/JP22KJ0728. This work was also supported by MEXT Promotion of Distinctive Joint Research Center Program Grant Number JPMXP0622717003. G. Tommei acknowledges the support from the Italian Space Agency (grant 2022-8-HH.0).

\bibliographystyle{ieeetr}
\bibliography{paperpile}

\end{document}